# SPEAKING STYLE AUTHENTICATION USING SUPRASEGMENTAL HIDDEN MARKOV MODELS

**Ismail Shahin**

Electrical and Computer Engineering Department, University of Sharjah
Sharjah, United Arab Emirates

## ABSTRACT

*The importance of speaking style authentication from human speech is gaining an increasing attention and concern from the engineering community. The importance comes from the demand to enhance both the naturalness and efficiency of spoken language human-machine interface. Our work in this research focuses on proposing, implementing, and testing speaker-dependent and text-dependent speaking style authentication (verification) systems that accept or reject the identity claim of a speaking style based on suprasegmental hidden Markov models (SPHMMs). Based on using SPHMMs, our results show that the average speaking style authentication performance is: 99%, 37%, 85%, 60%, 61%, 59%, 41%, 61%, and 57% belonging respectively to the speaking styles: neutral, shouted, slow, loud, soft, fast, angry, happy, and fearful.*

الخلاصة

أن أهمية التأكد أو التحقق من نمط أو أسلوب الحديث بين البشر تكتسب اهتماما متزايدا في الأوساط الهندسية. هذه الأهمية تأتي من الطلب على تحسين وزيادة كفاءة اللغة بين الإنسان والآلة. عملنا في هذا البحث يركز على اقتراح، وتنفيذ، واختبار التحقق من نمط الحديث لنظم تعتمد على نمط الكلام أو هوية نمط الكلام أو الحديث باستخدام طريقة **Suprasegmental Hidden Markov Models (SPHMMS)**. اعتمادا على هذه الطريقة، اظهرت نتائج بحثنا أن معدل تأكد نمط الكلام هو: **99%**، **37%**، **85%**، **60%**، **61%**، **59%**، **41%**، **61%**، و**57%**، منتمون على التوالي إلى أساليب الكلام: حيادي، صياح، بطيء، عالي، لين، سريع، غاضب، سعيد، وخائف.

## 1. INTRODUCTION

The goal of speech communication is to convey messages. Therefore, the message is the most important information embedded in the speech signal. But it is not the only information. Other kinds of information embedded in the speech signal include the identity of the speaker, the language spoken, the age of the speaker, the gender of the speaker, the accent of the speaker, the presence and type of speech pathologies, and the physical and emotional state of the speaker. Of these other kinds of information, the identity of the speaker has received most of the attention [1-4]. On the other hand, the language spoken, the age of the speaker, the gender of the speaker, the accent of the speaker, the presence and type of speech pathologies, and the physical and emotional state of the speaker have not received enough attention [5-7].

Speaking style recognition systems can operate in either an identification or authentication decision mode. In the identification mode, a speech sample from the unknown speaking style is analyzed and compared with models of known speaking styles. The unknown speaking style is identified as the speaking style whose model best matches the input speech sample. In the authentication decision mode, the objective is to decide whether a speaking style corresponds to a particular known speaking style or to some other unknown speaking styles. A speaking style known to the speaking style recognition system that is correctly claiming its identity is called a true speaking style and the speaking style unknown to the system that is posing as a known speaking style is called a false speaking style. A known speaking style is also referred to as a target speaking style, while a false speaking style is referred to as a background speaking style.

There are two types of errors in speaking style authentication systems: false acceptance, where a false speaking style is accepted, and false rejection, where a true speaking style is rejected.

Speaking style authentication systems typically operate in one of two input modes, text-dependent mode or text-independent mode. In the text-dependent mode, speaking styles must provide utterances of the same text for both training and testing (authentication) trials. In the text-independent mode, speaking styles are not constrained to provide specific texts in authentication trials. The process of speaking style recognition can be divided into two categories: "open set" and "closed set". In the "open set" category, a reference model for a test speaking style may not exist; whereas, in the "closed set" category, a reference model for a test speaking style must be available.

## 2. MOTIVATION

Speech is one of the important communication channels between a user and a computer and it can be used to recognize the speaking style status of the user. Speaking style recognition by speech is one of research fields for speaking style human-computer interaction or affective computing [8]. A major motivation comes from the desire to develop human-machine interface that is more adaptive and responsive to a user's behavior. The main task of intelligent human-machine interaction is to empower a computer with the affective computing ability so that a computer can recognize the speaking style of the user and then respond to the user in an affective method.

The applications of speaking style recognition systems are many. Speaking style recognition systems can be used in the applications of telecommunications, military field, and law enforcement. In telecommunications, speaking style recognition systems can be used to enhance the telephone-based speech recognition performance, route emergency call services for high precedence emergency calls, and assess a caller's speaking style for telephone response services. The integration of speech recognition technology is noticeable in many military voice communication and control applications. Such applications involve stressful environments such as aircraft cockpits and military peacekeeping [9]. Finally, speaking style recognition systems can be employed in forensic speech analysis by law enforcement to assess the state of telephone callers or as an aid in suspect interviews.

Researchers are incorporating emotional and stressful capabilities into speech synthesis programs, hoping to empower computers that can communicate emotionally and stressfully with users through expressive vocal signals such as laughter, sighing, or sad tones of voice. IBM is set to release a new Expressive Text-to-Speech Engine for commercial use that will deliver spoken information in the appropriate tone (sadness, happiness, frustration, etc…). AT&T Lab is developing the opposite technology, software that can detect users' emotional and stressful state; voice-response systems equipped with this software would be able to prioritize calls according to the person's state of agitation, for example [10].

In the last four decades, all the attentions and concerns have been focused on speech recognition and speaker recognition areas. Researchers focus their work on these two areas under the neutral and stressful speaking styles [1,2,11,12]. On the other hand, the area of speaking style recognition does not receive enough attention [6,7,13,14]. Dealing with the area of speaking style recognition is one of the latest challenges in speech technologies. Three different aspects can be easily identified in the area of speaking styles: speech recognition in the presence of speaking styles, synthesis of speaking styles, and speaking style recognition [13]. Lee and Narayanan focused their work on recognizing emotions from spoken language [6]. In their work, Lee and Narayanan used a combination of three sources of information for emotion recognition. The three sources are: acoustic, lexical, and discourse. Li and Zhao worked on recognizing emotions in speech using short-term and long-term features [15]. Wu *et al.* focused their work on studying the influence of emotion on the performance of a GMM-UBM based speaker verification system [16]. In their work, they proposed an emotion-dependent score normalization for speaker verification on emotional speech.

Our work in this paper focuses on the process of accepting or rejecting the identity claim of a speaking style of speaker-dependent and text-dependent speaking style authentication systems based on SPHMMs. This is the first known work that focuses on speaking style authentication systems based on SPHMMs. Our speaking styles in this work are: neutral, shouted, slow, loud, soft, fast, angry, happy, and fearful.

Speaker-dependent and text-dependent speaking style authentication systems are currently being developed with the aim of enhancing the quality of human-computer interaction. The authentication of the speaking style user state (speaker-dependent) seems often essential to verify the true meaning (text-dependent) of what has been said or to adapt the computer's behavior to the user's speaking styles.

This paper is organized as follows. Section 3 explains briefly speaking styles. Section 4 discusses suprasegmental hidden Markov models. Section 5 describes the speech database used. Section 6 discusses the algorithm of speaking style authentication systems based on each of hidden Markov models (HMMs) and SPHMMs. Results and discussion appear in Section 7. Concluding remarks are listed in Section 8.

## 3.  SPEAKING STYLES

Neutral speaking style can be defined as the speaking style in which speech is produced assuming that the speaker is in a "quiet room" with no task obligations. Speaking styles other than the neutral speaking style are defined as speaking styles that cause a speaker to vary his/her production of speech from the neutral speaking style.

Researchers in the areas of speech recognition, speaker recognition, and speaking style recognition use speech database to test their proposed algorithm. There are many speech databases that are available to researchers in the three areas. Ververidis and Kotropoulos reviewed 32 speaking style and emotional speech databases [17]. Each database consists of a corpus of human speech uttered under different speaking styles and emotional states.

One of the most famous speech databases is the SUSAS (Speech Under Simulated and Actual Stress) database [18]. This database was established to conduct research studies for analysis and algorithm formulation of speech recognition and speaker recognition in stressful and noisy environment. This database can also be used in the area of speaking style recognition. Hansen, Cummings, and Clements used the SUSAS speech database in which eight talking conditions are used to simulate speech produced under real stressful talking conditions and three real talking conditions [11,12,19].

## 4. SUPRASEGMENTAL HIDDEN MARKOV MODELS

HMMs have been successfully used in the fields of speech recognition and speaker recognition applications in the last three decades. HMMs have become one of the most successful and broadly used modeling techniques in the two fields [3,20,21]. Remarkably, robust models of speech events can be obtained with only small amounts of specification or information accompanying training utterances. In HMMs, formulation the speech signal is considered as a sequence of Markov states representing transitions from one speech event to another. The Markov states themselves are "hidden" but are indirectly observable from the sequences of spectral feature vectors.

HMMs are powerful models in optimizing the parameters that are used in modeling speech signals. This optimization decreases the computational complexity in the decoding procedure and improves the recognition accuracy [21]. HMMs use Markov chain to model the changing statistical characteristics that exist in the actual observations of speech signals. HMMs are double stochastic processes where there is an unobservable Markov chain defined by a state transition matrix, and where each state of the Markov chain is associated with either a discrete output probability distribution (discrete HMMs) or a continuous output probability density function (continuous HMMs) [21].

A suprasegmental is a vocal effect that extends over more than one sound segment in an utterance, such as pitch, stress, or juncture pattern. Suprasegmental is often used for tone, vowel length, and features like nasalization and aspiration. Stress, tone, intonation, length, and organization of segments into syllables are usually considered suprasegmental properties.

The ability and capability of SPHMMs appear clearly in summarizing several states with HMMs into what is termed a suprasegmental state. Suprasegmental states are able and capable of looking at the observation sequence through a larger window to capture prosodic properties. Such states allow observations at rates appropriate for the situation of speaking style modeling. For example, prosodic information can not be observed at a rate that is used to model the acoustic features of speech signals. Fundamental frequency, intensity, and duration of speech signals are the main acoustic parameters that describe prosody [22].

In linguistics, prosody is the intonation, rhythm, and lexical stress in speech. The prosodic features of a unit of speech are called suprasegmental features because they affect all the segments of the unit. These features are manifested, among other things, as syllable length, tone, and stress. Prosodic information can be applied to phrases, words, syllables, and phones which of course can not be observed in the time frame in which acoustic information can be observed. Hence, prosodic events are modeled using suprasegmental states, while acoustic events are modeled using conventional hidden Markov states.

Lea proposed the use of prosodic information in Automatic Speech Understanding (ASU) systems [23]. Seventeen years later, a German speech-to-speech translation system called VERBMOBIL was completed as the world wide first complete speech understanding system, where prosody was really used. It was shown by the VERBMOBIL system that the used and implemented prosody yielded drastic performance improvement.

The basic structure of SPHMMs is given (as an example) in Fig. 1. $q_1, q_2, ...,q_6$ are hidden Markov states. $p_1$ is a suprasegmental state (e.g. phone) that is composed of $q_1, q_2$, and $q_3$. $p_2$ is a suprasegmental state (e.g. phone) that is composed of $q_4, q_5$, and $q_6$. $p_3$ is a suprasegmental state (e.g. syllable) that is made up of $p_1$ and $p_2$. $a_{ij}$ is the transition probability between the $i$th hidden Markov state and the $j$th hidden Markov state. $b_{ij}$ is the transition probability between the $i$th suprasegmental state and the $j$th suprasegmental state.

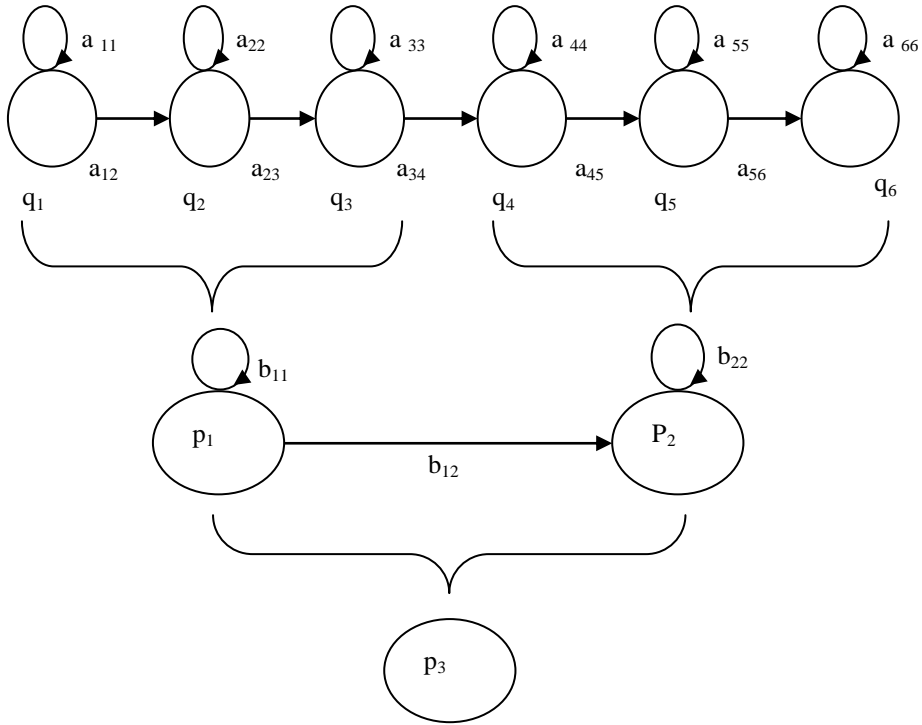

**Fig. 1.** Basic structure of suprasegmental hidden Markov models

## 5. SPEECH DATABASE

Our speaking style speech corpus was collected from 20 untrained healthy adult native speakers of American English (10 males and 10 females). Untrained speakers were selected to avoid exaggerated expressions. Each speaker uttered 8 sentences where each sentence was uttered 9 times (9 utterances or tokens per sentence) under each of the neutral, shouted, slow, loud, soft, fast, angry, happy, and fearful speaking styles. In addition, we asked the 20 speakers to utter the 8 sentences for a sad speaking style.

The 8 sentences were unbiased towards any speaking style (no correlation between any sentence and any speaking style). These sentences were:

1) *He works five days a week.*
2) *The sun is shining.*
3) *The weather is fair.*
4) *The students study hard.*
5) *Assistant professors are looking for promotion.*
6) *University of Sharjah.*
7) *Electrical and Computer Engineering Department.*
8) *He has two sons and two daughters.*

Our speech database was recorded in a clean environment that was not affected by background noise. The speech database was captured by a speech acquisition board using a 16-bit linear coding A/D converter and sampled at a sampling rate of 16 kHz. Our database was a wideband 16-bit per sample linear data. The signal samples were segmented into frames of 16 ms each with 9 ms overlap between consecutive frames. The total number of frames to be processed depends on the length of the utterance. The speech signals were then applied to the Hamming window. Next, $16^{th}$ order linear prediction coefficients (LPCs) were extracted from each frame by the autocorrelation method. The $16^{th}$ order LPCs were then transformed into $16^{th}$ order Linear Prediction Cepstral Coefficients (LPCCs).

The LPCC feature analysis was used to form the observation vectors in each of HMMs and SPHMMs. In HMMs, the number of states, $N$, was 5 and the number of mixture components, $M$, was 5 per state. In SPHMMs, the number of suprasegmental states was 2 ($p_1$ and $p_2$, where $p_1$ was composed of 3 hidden Markov states: $q_1, q_2,$ and $q_3$, while $p_2$ was composed of 2 hidden Markov states: $q_4$ and $q_5$) and the number of mixture components was 5 per state. A continuous mixture observation density was selected for each of HMMs and SPHMMs.

## 6. THE ALGORITHM OF SPEAKING STYLE AUTHENTICATION SYSTEMS BASED ON EACH OF HMMS AND SPHMMS

The general overview of text-dependent speaking style authentication system is shown in Fig. 2.

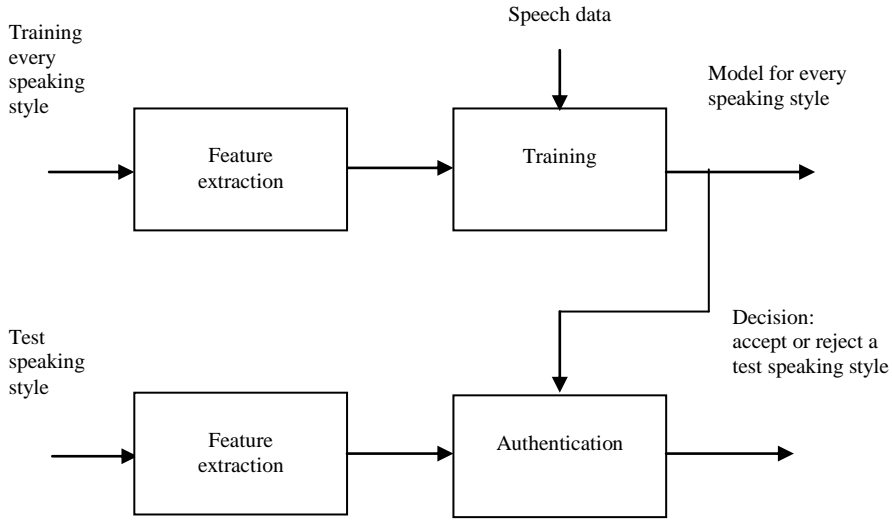

**Fig. 2.** General overview of text-dependent speaking style authentication system

Speaking style authentication problem in each of HMMs and SPHMMs requires making a binary decision based on two hypotheses. Hypothesis $H_0$ if a test speaking style belongs to a true speaking style (claimant) or hypothesis $H_1$ if a test speaking style comes from a false speaking style (imposter).

### Talking Condition Authentication Based on HMMs

In the training session of the conventional HMMs, one reference model per speaker per sentence per speaking style was derived using 5 of the 9 utterances. The total number of utterances in this session was 7200. Training of models in this session used the forward-backward algorithm.

In the testing (authentication) session, all the nine speaking styles were used in addition to the sad speaking style since our speech database was an open set. In this session, each one of the 10 speaking styles used 4 of the 9 utterances per the same speaker per the same sentence (5 of the 9 utterances of the sad speaking style were not used). The total number of utterances in this session was 6400. The authentication process in this session used the Viterbi decoding algorithm.

Based on HMMs, the log-likelihood ratio can be defined as [24],

$$\Lambda_{HMMs}(O) = log\left[P(O|\lambda_C)\right] - log\left[P(O|\lambda_{\overline{C}})\right] \tag{1}$$

where,

$\Lambda_{HMMs}$ : is the log-likelihood ratio based on HMMs.

$O$: is the observation vector or sequence that belongs to the test speaking style.

$P(O|\lambda_c)$: is the probability that a test speaking style comes from a true speaking style based on HMMs.

$\lambda_C$: is the HMMs claimant speaking style model.

$P(O|\lambda_{\overline{c}})$: is the probability that a test speaking style comes from a false speaking style based on HMMs.

$\lambda_{\overline{C}}$ : is the HMMs imposter speaking style model.

The last step in the authentication process is to compare the HMMs threshold ($\theta_{HMMs}$) with the log-likelihood ratio in order to accept or reject the test speaking style that belongs to the claimed speaking style, i.e.,

$$\text{Accept a claimed speaking style if } \Lambda_{HMMs}(O) \geq \theta_{HMMs}$$

$$\text{Reject a claimed speaking style if } \Lambda_{HMMs}(O) < \theta_{HMMs}$$

## Speaking Style Authentication Based on SPHMMs

In the training session of SPHMMs, the training process was very similar to the training process of conventional HMMs. In the training session of SPHMMs, suprasegmental models were trained on top of acoustic models. In this session, one reference model per speaker per sentence per speaking style was derived using 5 of the 9 utterances. The testing session of SPHMMs is the same as the testing session of HMMs.

Based on SPHMMs, the log-likelihood ratio can be defined as,

$$\Lambda_{SPHMMs}(O) = log\left[P\left(O|\Psi_C\right)\right] - log\left[P\left(O\,\middle|\,\Psi_{\overline{C}}\right)\right] \tag{2}$$

where,

$\Lambda_{SPHMMs}$ : is the log-likelihood ratio based on SPHMMs.

$P\left(O|\Psi_C\right)$: is the probability that a test speaking style comes from a true speaking style based on SPHMMs.

$\Psi_C$: is the SPHMMs claimant speaking style model.

$P\left(O|\Psi_{\overline{C}}\right)$: is the probability that a test speaking style comes from a false speaking style based on SPHMMs.

$\Psi_{\overline{C}}$ : is the SPHMMs imposter speaking style model.

The last step in the authentication process in SPHMMs is to compare the SPHMMs threshold ($\theta_{SPHMMs}$) with the log-likelihood ratio in order to accept or reject a test speaking style that belongs to the claimed speaking style, i.e.,

Accept a claimed speaking style if $\Lambda_{SPHMMs}(O) \geq \theta_{SPHMMs}$

Reject a claimed speaking style if $\Lambda_{SPHMMs}(O) < \theta_{SPHMMs}$

In equations (1) and (2), the log-likelihood ratio depends on the probability that a test speaking style comes from a true speaking style and the probability that a test speaking style comes from a false speaking style based on HMMs and SPHMMs, respectively. There are three scenarios to determine the log-likelihood ratio. These three scenarios are:

**Scenario 1:** The probability that a test speaking style utterance does not come from the claimant speaking style is determined from a set of imposter speaking style models: $\lambda_{\overline{C}_1}, \lambda_{\overline{C}_2}, ...., \lambda_{\overline{C}_K}$, where $K$ is the number of imposter speaking style models. In this scenario, the imposter speaking style model, $\overline{C}$, is obtained from these individual models.

**Scenario 2:** The claimant log-probability score is computed first and then the maximum of the imposter scores over individual imposter models is subtracted from the claimant log-probability score for the HMMs and SPHMMs as given in the following two formulae, respectively,

$$\Lambda_{\text{HMMs}}(O) = log\left[P(O|\lambda_C)\right] - \max_i \; log\left[P(O|\lambda_{\overline{C}_i})\right] \tag{3}$$

where $\overline{C}_i$ is the $i$th individual imposter model.

$$\Lambda_{\text{SPHMMs}}(O) = log\left[P(O|\Psi_C)\right] - \max_i \; log\left[P(O|\Psi_{\overline{C}_i})\right] \tag{4}$$

**Scenario 3:** Simply, the claimant log-probability score, $log\left\{P(O|\lambda_C)\right\}$ or $log\left\{P(O|\Psi_C)\right\}$, is used only without any form of imposter score. In this work, we adopted this scenario for simplicity. Therefore, the threshold is sensitive to the probability that a test speaking style comes from a true speaking style.

To assign an appropriate value of threshold to our authentication system based on each of HMMs and SPHMMs, this value of threshold should be able to tolerate trial-to-trial variations and at the same time yields a desired level of performance. A tight value of threshold makes it difficult for false speaking styles to be falsely accepted but at the expense of falsely rejecting true speaking styles. On the other hand, a loose value of threshold enables true speaking styles to be accepted constantly at the expense of falsely accepting false speaking styles. In order to set an appropriate value of threshold that satisfies a desired level of a true speaking style rejection and a false speaking style acceptance, it is necessary to know the distribution of true speaking style and false speaking style scores. A reasonable procedure for setting a value of threshold is to assign a loose initial value of threshold and then let it adapt by setting it to the average of recent trial scores. This loose value of threshold provides inadequate protection against false speaking style attempts.

In each of HMMs and SPHMMs, an input utterance from a test speaking style was analyzed to extract speaking style characteristic features. The measured features were compared with prototype features obtained from known speaking style models. In the speaking style identification mode, the comparison was carried out with every speaking style model (there are *n* comparisons or tests for *n* speaking styles). In the speaking style authentication mode, the comparison was carried out only with the model corresponding to the claimed speaking style identity (there is one comparison or test for a given unknown speaking style). In the speaking style authentication systems, an identity claim is made by or asserted for the unknown speaking style. The speech sample of the unknown speaking style is compared with the model for the speaking style whose identity is claimed. In such systems, there are two decision alternatives, accept or reject the identity claim, regardless of the size of the speaking styles.

## 7. RESULTS AND DISCUSSION

Our work in this research focused on verifying speaker-dependent and text-dependent speaking styles from spoken English language based on SPHMMs. We compare our results based on SPHMMs as shown in Table 1 with that based on HMMs as shown in Table 2. It is evident from Table 1 and Table 2 that SPHMMs are superior models over HMMs for speaking style authentication systems. The average speaking style authentication performance based on HMMs is 57.7%; on the other hand, the average speaking style authentication performance based on SPHMMs is 62.2%. The improvement rate of using SPHMMs over HMMs is 7.8%. This may be attributed to the following reasons:

1. SPHMMs are suitable and sufficient models to integrate observations from different modalities because such models allow for observations at a rate appropriate for each modality. On the other hand, HMMs are inappropriate and insufficient models to integrate observations from different modalities because such models are not capable for observations at a rate appropriate for each modality. This is because suprasegmental states allow observations at rates appropriate for the phenomena they are intended to model. This property of suprasegmental states is not possessed by hidden Markov states. In a work done by Polzin and Waibel, suprasegmental hidden Markov models yield better emotional detection accuracy than hidden Markov models [25].

2. Prosodic information adds more discriminative power to the speaking style authentication systems than the acoustic information.

**Table 1:** Speaking style authentication performance based on SPHMMs

| Speaking style | Males | | Females | | Average | |
|---|---|---|---|---|---|---|
| | $H_0$ | $H_1$ | $H_0$ | $H_1$ | $H_0$ | $H_1$ |
| Neutral | 99% | 1% | 99% | 1% | 99% | 1% |
| Shouted | 36% | 22% | 38% | 20% | 37% | 21% |
| Slow | 84% | 12% | 86% | 12% | 85% | 12% |
| Loud | 60% | 17% | 60% | 15% | 60% | 16% |
| Soft | 60% | 18% | 62% | 18% | 61% | 18% |
| Fast | 60% | 18% | 58% | 18% | 59% | 18% |
| Angry | 40% | 21% | 42% | 23% | 41% | 22% |
| Happy | 60% | 18% | 62% | 18% | 61% | 18% |
| Fearful | 56% | 19% | 58% | 19% | 57% | 19% |

**Table 2:** Speaking style authentication performance based on HMMs

| Speaking style | Males | | Females | | Average | |
|---|---|---|---|---|---|---|
| | $H_0$ | $H_1$ | $H_0$ | $H_1$ | $H_0$ | $H_1$ |
| Neutral | 99% | 1% | 99% | 1% | 99% | 1% |
| Shouted | 30% | 25% | 34% | 23% | 32% | 24% |
| Slow | 78% | 15% | 82% | 13% | 80% | 14% |
| Loud | 54% | 20% | 56% | 18% | 55% | 19% |
| Soft | 56% | 19% | 58% | 19% | 57% | 19% |
| Fast | 50% | 21% | 56% | 19% | 53% | 20% |
| Angry | 38% | 24% | 36% | 26% | 37% | 25% |
| Happy | 55% | 20% | 55% | 18% | 55% | 19% |
| Fearful | 52% | 22% | 50% | 20% | 51% | 21% |

Table 1 shows evidently that the neutral speaking style authentication performance of a test speaking style that belongs to a true speaking style ($H_0$) is 99% and the neutral speaking style authentication performance of a test speaking style that comes from a false speaking style ($H_1$) is 1%. The improvement rate of using SPHMMs over HMMs under such a speaking style is 0%. This means that both SPHMMs and HMMs are efficient and sufficient models under the neutral speaking style.

Table 3 and Table 4 show confusion matrices based on SPHMMs and HMMs, respectively. The two tables represent the percentage of confusion of a test speaking style with the other speaking styles based on SPHMMs and HMMs, respectively. These two tables show the following:

**Table 3:** Confusion matrix based on SPHMMs

| | Percentage of confusion of a test speaking style with the other speaking styles | | | | | | | | |
|---|---|---|---|---|---|---|---|---|---|
| Model | Neutral | Shouted | Slow | Loud | Soft | Fast | Angry | Happy | Fearful |
| Neutral | 99% | 0% | 7% | 3% | 6% | 3% | 2% | 2% | 4% |
| Shouted | 0% | 36% | 2% | 16% | 2% | 9% | 26% | 3% | 6% |
| Slow | 1% | 0% | 83% | 0% | 7% | 2% | 1% | 3% | 13% |
| Loud | 0% | 28% | 2% | 54% | 2% | 10% | 13% | 10% | 6% |
| Soft | 0% | 0% | 2% | 0% | 65% | 2% | 0% | 2% | 8% |
| Fast | 0% | 6% | 0% | 7% | 2% | 58% | 10% | 13% | 3% |
| Angry | 0% | 23% | 0% | 11% | 1% | 9% | 44% | 0% | 8% |
| Happy | 0% | 0% | 2% | 9% | 6% | 7% | 0% | 64% | 2% |
| Fearful | 0% | 7% | 2% | 0% | 9% | 0% | 4% | 3% | 50% |

1. The most easily recognizable speaking style is neutral. Hence, the performance of verifying the neutral speaking style is the highest compared to verifying the other speaking styles as shown in Table 1 and Table 2.

2. The second most easily recognizable speaking style is slow. Therefore, the performance of verifying the slow speaking style is the second highest compared to the performance of the other speaking styles as shown in Table 1 and Table 2.

3. The least easily recognizable speaking style is shouted. Consequently, the least speaking style verification performance occurs when the test speaking style is shouted as shown clearly in Table 1 and Table 2.

**Table 4:** Confusion matrix based on HMMs

| | Percentage of confusion of a test speaking style with the other speaking styles | | | | | | | | |
|---------|---------|---------|------|------|------|------|-------|-------|---------|
| Model | Neutral | Shouted | Slow | Loud | Soft | Fast | Angry | Happy | Fearful |
| Neutral | 98% | 0% | 10% | 3% | 8% | 3% | 2% | 2% | 4% |
| Shouted | 0% | 29% | 2% | 20% | 2% | 12% | 31% | 3% | 8% |
| Slow | 1% | 0% | 78% | 0% | 10% | 2% | 1% | 3% | 16% |
| Loud | 0% | 33% | 2% | 48% | 2% | 12% | 15% | 14% | 6% |
| Soft | 1% | 0% | 2% | 0% | 59% | 2% | 0% | 2% | 9% |
| Fast | 0% | 6% | 0% | 7% | 2% | 51% | 10% | 17% | 3% |
| Angry | 0% | 25% | 0% | 12% | 1% | 11% | 37% | 0% | 8% |
| Happy | 0% | 0% | 2% | 10% | 6% | 7% | 0% | 56% | 2% |
| Fearful | 0% | 7% | 4% | 0% | 10% | 0% | 4% | 3% | 44% |

4. Column 3 of Table 3, for example, shows that based on SPHMMs 0% of the utterances that were portrayed as shouted speaking style were evaluated as neutral, slow, soft, and happy speaking styles (the proximity between the shouted and each of the neutral, slow, soft, and happy is zero), 28% of the utterances that were portrayed as shouted speaking style were evaluated as loud speaking style. This column shows that the shouted speaking style has the highest confusion percentage with the loud speaking style (28%) and with the angry speaking style (23%). Therefore, the shouted speaking style is highly confusable with each of the loud and angry speaking styles.

5. Table 3 and Table 4 show evidently why SPHMMs outperform HMMs for speaking style authentication systems.

It is evident from Table 1 that the least speaking style authentication performance occurs under the shouted speaking style. The reason for this sharply degraded authentication performance is that the shouted speaking style is considered as the most stressful speaking style [3]. It is apparent from Table 1 and Table 2 that HMMs yield lower authentication performance than SPHMMs under such a speaking style. In fact, HMMs are inefficient and insufficient models under the shouted speaking style because the changes in the statistical characteristics that exist in the actual observations of the shouted speaking style are greater than those of the neutral speaking style. This sharply degraded authentication performance comes from the fact that the log-likelihood ratio is much less than the threshold. The reason of this low log-likelihood ratio is that the probability that a test speaking style comes from a true speaking style and the probability that a test speaking style comes from a false speaking style are very close. It was reported in many other studies that HMMs did not perform well under such a speaking style [1-3]. The improvement rate of using SPHMMs over HMMs under such a speaking style is very significant (15.6%). Table 4 shows that the shouted speaking style is highly confusable with each of the loud and angry speaking styles. This high proximity between the shouted speaking style and each of the loud and angry speaking styles explains the degraded shouted speaking style verification performance.

Table 1 shows that the slow speaking style authentication performance is quite high compared to the other speaking styles. This is because the slow speaking style is rarely confusable with other speaking styles as shown clearly in Table 3 and Table 4. Table 2 shows that HMMs are convenient models for such a speaking style. This is because the talking rate under the slow speaking style is low. Low talking rate results in a sufficient number of frames per state in HMMs. Therefore, HMMs represent slow speaking style speech signals efficiently and conveniently. Table 1 and Table 2 show that SPHMMs noticeably enhance the authentication performance compared to HMMs.

The loud speaking style authentication performance is quite low. Comparing the loud speaking style authentication performance with that of the shouted speaking style, it is clear that the loud speaking style authentication performance is almost twice as that for the shouted speaking style. This is because the shouted speaking style consists of two components: loud speaking style and noise [26]. The improvement rate of using SPHMMs over HMMs under such a speaking style is significant (9.1%). Table 3 and Table 4 show that this speaking style is highly confusable with the shouted speaking style.

The soft speaking style authentication performance is low. This is because the soft speaking style approaches breathy or whispering speaking style which means that the speaking style authentication system will not be able to verify correctly when a test speaking style is soft. It is difficult for speaking style authentication system to verify correctly breathy or whispering speech signals.

Table 1 clearly shows that the fast speaking style authentication performance based on SPHMMs is improved remarkably compared to that based on HMMs; the improvement rate of using SPHMMs over HMMs is 11%. This is because suprasegmental states have the ability and capability to look at the observation sequence through a larger window. Such states allow observations at rates suitable for the situation of modeling.

The angry speaking style authentication performance is very low. The authentication performance for this speaking style is the second lowest authentication performance based on each of SPHMMs and HMMs. The authentication performance of each the shouted and angry speaking styles is close to each other since the shouted speaking style can not be entirely separated from the angry speaking style in our real life [1]. It is obvious from the two confusion matrices that the angry speaking style is highly confusable with the shouted speaking style.

Based on SPHMMs, the happy speaking style authentication performance is quite low. This is because when people talk under joy and happiness, they do not usually talk clearly. Consequently, there will be confusion and ambiguity in their speech. The confusion happens between the happy speaking style and each of the loud and fast speaking styles. The improvement rate of using SPHMMs over HMMs under such a speaking style is significant (10.9%).

The fearful speaking style authentication performance is worse than that of the happy speaking style. This is because when people talk under the fearful and frightened mode, they produce speech with more ambiguity and less clarity than that under the happy mode. The fearful speaking style has a remarkable confusion with each of the slow, soft, and angry speaking styles. The improvement rate of using SPHMMs over HMMs under such a speaking style is significant (11.8%).

It is noticeable from Table 1 that for all of the speaking styles (except for the fast speaking style), the speaking style authentication performance based on SPHMMs for the female speakers is higher than that for the male speakers. This is because as shown from Table 5 and Table 6 that the confusion percentage of a female test speaking style with the other female speaking styles is less than that for a male test speaking style. It seems that SPHMMs add more discriminative power to the female speaking styles than that to the male speaking styles.

**Table 5:** Confusion matrix for male speakers based on SPHMMs

| Model | Percentage of confusion of a test speaking style with the other speaking styles | | | | | | | | |
|---|---|---|---|---|---|---|---|---|---|
| | Neutral | Shouted | Slow | Loud | Soft | Fast | Angry | Happy | Fearful |
| Neutral | 99% | 0% | 9% | 3% | 6% | 3% | 2% | 2% | 4% |
| Shouted | 0% | 34% | 2% | 18% | 2% | 9% | 29% | 3% | 7% |
| Slow | 1% | 0% | 80% | 0% | 8% | 2% | 1% | 3% | 14% |
| Loud | 0% | 29% | 2% | 52% | 2% | 10% | 13% | 11% | 6% |
| Soft | 0% | 0% | 2% | 0% | 62% | 2% | 0% | 2% | 8% |
| Fast | 0% | 6% | 0% | 7% | 2% | 58% | 10% | 14% | 3% |
| Angry | 0% | 24% | 0% | 11% | 1% | 9% | 41% | 0% | 8% |
| Happy | 0% | 0% | 2% | 9% | 6% | 7% | 0% | 62% | 2% |
| Fearful | 0% | 7% | 3% | 0% | 11% | 0% | 4% | 3% | 48% |

A new method called multi-speaker training method has been proposed and implemented to enhance speaking style authentication performance. Multi-speaker training method requires a speaker to train an authentication system using speaking styles spoken by different speakers instead of using speaking styles all spoken by the same speaker.

**Table 6:** Confusion matrix for female speakers based on SPHMMs

| Model | Percentage of confusion of a test speaking style with the other speaking styles | | | | | | | | |
|---|---|---|---|---|---|---|---|---|---|
| | Neutral | Shouted | Slow | Loud | Soft | Fast | Angry | Happy | Fearful |
| Neutral | 99% | 0% | 6% | 3% | 6% | 3% | 2% | 2% | 4% |
| Shouted | 0% | 37% | 2% | 16% | 2% | 8% | 24% | 3% | 6% |
| Slow | 1% | 0% | 84% | 0% | 7% | 2% | 1% | 3% | 12% |
| Loud | 0% | 27% | 2% | 54% | 2% | 9% | 13% | 9% | 6% |
| Soft | 0% | 0% | 2% | 0% | 66% | 2% | 0% | 2% | 7% |
| Fast | 0% | 6% | 0% | 7% | 2% | 60% | 10% | 13% | 3% |
| Angry | 0% | 23% | 0% | 11% | 1% | 9% | 46% | 0% | 8% |
| Happy | 0% | 0% | 2% | 9% | 6% | 7% | 0% | 65% | 2% |
| Fearful | 0% | 7% | 2% | 0% | 8% | 0% | 4% | 3% | 52% |

Two different experiments on multi-speaker trained speaking style authentication systems have been performed. One experiment trained based on SPHMMs and the other one trained based on HMMs. In each experiment, for each speaking style, a multi-speaker trained model added one utterance from each speaker per sentence to the previous training set taking into considerations that none of the test utterances are used for training. Table 7 and Table 8 summarize the results of speaking style authentication performance using the multi-speaker training method based on SPHMMs and HMMs, respectively.

Comparing Table 7 with Table 8, it is evident that the speaking style authentication performance based on the multi-speaker SPHMMs training method has been significantly improved compared to that based on the multi-speaker HMMs training method. This significant enhancement of using the multi-speaker SPHMMs training method over the multi-speaker HMMs training method shows evidently that SPHMMs are superior models over HMMs for speaking style authentication systems.

**Table 7:** Speaking style authentication performance using multi-speaker SPHMMs training method

| Speaking style | Males | | Females | | Average | |
|---|---|---|---|---|---|---|
| | $H_0$ | $H_1$ | $H_0$ | $H_1$ | $H_0$ | $H_1$ |
| Neutral | 99% | 1% | 99% | 1% | 99% | 1% |
| Shouted | 38% | 17% | 40% | 17% | 39% | 17% |
| Slow | 85% | 10% | 85% | 10% | 85% | 10% |
| Loud | 63% | 15% | 61% | 15% | 62% | 15% |
| Soft | 59% | 14% | 61% | 14% | 60% | 14% |
| Fast | 59% | 15% | 61% | 15% | 60% | 15% |
| Angry | 44% | 18% | 40% | 18% | 42% | 18% |
| Happy | 61% | 13% | 63% | 15% | 62% | 14% |
| Fearful | 57% | 15% | 57% | 15% | 57% | 15% |

# 8. CONCLUDING REMARKS

Our work in this research focused on enhancing speaker-dependent and text-dependent speaking style authentication systems based on SPHMMs. Generally speaking, speaking style authentication systems are not straightforward systems. The field of human speaking styles is an enormously complicated field of study. This may be attributed to a number of considerations:

**Table 8:** Speaking style authentication performance using multi-speaker HMMs training method

| Speaking style | Males | | Females | | Average | |
|---|---|---|---|---|---|---|
| | $H_0$ | $H_1$ | $H_0$ | $H_1$ | $H_0$ | $H_1$ |
| Neutral | 99% | 1% | 99% | 1% | 99% | 1% |
| Shouted | 34% | 18% | 36% | 22% | 35% | 20% |
| Slow | 83% | 12% | 83% | 10% | 83% | 11% |
| Loud | 60% | 17% | 58% | 15% | 59% | 16% |
| Soft | 59% | 14% | 59% | 16% | 59% | 15% |
| Fast | 54% | 16% | 58% | 16% | 56% | 16% |
| Angry | 40% | 20% | 38% | 20% | 39% | 20% |
| Happy | 59% | 14% | 59% | 16% | 59% | 15% |
| Fearful | 55% | 16% | 55% | 16% | 55% | 16% |

1) Voice quality is intrasegmental since it is dependent on each individual vocal tract.

2) Speaking style authentication systems depend on many areas such as signal processing and analysis techniques, psychology, physiology, and linguistics. Therefore, it is recommended to include these areas for any future work that is related to the speaking style recognition.

3) There is a lack of complete understanding and knowledge of speaking styles in human minds, including a lack of agreement among psychologists.

4) Speakers usually use certain words more frequently in expressing their speaking styles since they have learned the connection between certain words and their corresponding speaking styles. This issue is studied thoroughly in the field of psychology [27]. For example, people under the happy mode frequently use the words: *yes*, *agree*, *correct*, *good*, etc…; on the other hand, people under the angry mode use frequently the words: *no*, *disagree*, *wrong*, *bad*, etc…